# Direct Optical-Structure Correlation in Atomically Thin Dichalcogenides and Heterostructures

*Akshay Singh, Hae Yeon Lee, Silvija Gradečak\**

Department of Materials Science and Engineering, Massachusetts Institute of Technology, 77 Massachusetts Ave, Cambridge, MA 02141, USA

**Corresponding Author:** Silvija Gradecak ; email: gradecak@mit.edu

ABSTRACT: Atomically thin transition metal dichalcogenides (TMDs) have distinct opto-electronic properties including enhanced luminescence and high on-off current ratios, which can be further modulated by making more complex TMD heterostructures. However, resolution limits of conventional optical methods do not allow for direct nanoscale optical–structural correlation measurements in these materials, particularly of buried interfaces in TMD heterostructures. Here we use, for the first time, electron beam induced cathodoluminescence in a scanning transmission electron microscope (CL-STEM) to measure optical properties of monolayer TMDs ($WS_2$, $MoS_2$ and WSSe alloy) encapsulated between layers of hBN. We observe dark areas resulting from localized (~ 100 nm) imperfect interfaces and monolayer folding, which shows that the intimate contact between layers in this application-relevant heterostructure is required for proper inter layer coupling. We also realize a suitable imaging method that minimizes electron-beam induced changes and provides measurement of intrinsic properties. To overcome the limitation of small electron interaction volume in TMD monolayer (and hence low photon yield), we find that encapsulation of TMD monolayers with hBN and subsequent annealing is important. CL-STEM offers to be a powerful method to directly measure structure-optical correspondence in lateral or vertical heterostructures and alloys.

TEXT: Transition metal dichalcogenides (TMDs) in the monolayer limit show unique optical properties including enhanced photoluminescence (PL) and spin-valley coupling [1, 2]. The fundamental optical transition is excitonic (Coulomb bound electron-hole pair), which persists at room temperatures as a consequence of large ~200-400 meV binding energies [3, 4]. Functionality beyond monolayers can be enhanced when using mixed (alloys) or separated (lateral or vertical heterostructures) multi-phase TMDs [5, 6]. These multi-phase materials have a wide range of applications including visible light photodetection, photovoltaics, and optical memories [7, 8]. Multi-phase materials have been synthesized as well as mechanically stacked, with the interface quality and dielectric environment emerging as important characteristics to control layer coupling and opto-electronic properties [5, 9, 10].

Encapsulation between layers of a dielectric has shown to be important in measuring and controlling intrinsic opto-electronic properties of TMDs and other two-dimensional materials [11, 12]. Using this approach, enhanced opto-electronic properties such as reduced substrate doping, higher carrier mobilities enabling observation of quantum phenomena, and line-width narrowing have been observed [9, 13, 14]. Further, intrinsic coupling behavior and improvements in opto-electronic properties are observed when layers of a multi-phase van der Waals (vdW) material are in intimate contact. However, effects of the interface properties in these stacked structures have only been indirectly probed. For example, lower transistor mobility or lower inter-layer coupling have been attributed to imperfect interfaces [10, 15], but these earlier works are either non-local measurements (electrical) or they suffer from insufficient spatial resolution (optical). Thus, only limited information on effect of nanoscale inhomogeneities on opto-electronic properties has been uncovered for vdW heterostructures, with no direct optical-structure correlation.

Transmission electron microscopy (TEM) has been used extensively to investigate structure of multi-phase TMDs occurring naturally or grown by synthetic processes [11, 16]. Limited cathodoluminescence (CL) measurements using scanning electron microscopy (SEM) have provided interesting insights into luminescence of layered materials changes due to sulfur vacancies, encapsulation with dielectric layers, and coupling with other materials [17-19]. However, all of these measurements have relied on SEM, which offers limited spatial resolution compared to scanning transmission electron microscopy (STEM) due to beam broadening and larger probe sizes, and limited compositional contrast due to reliance on secondary electrons (see electronic supplementary material (ESM) and refs [20, 21]). We and others have previously shown that by combining CL and STEM, it is possible to directly correlate structural and optical properties with nanoscale resolution in a range of nanostructured materials [20, 22-25]. CL-STEM offers significantly higher spatial resolution – limited fundamentally by the probe size and carrier diffusion lengths – compared to other optical techniques. Thus far, CL-STEM measurements on layered materials have only been performed on thick few-layer samples [26, 27], but it remains unclear if the interaction of high-energy electrons in STEM with a monolayer material is sufficient to generate luminescence signal for quantitative CL analysis.

In this work, we demonstrate quantitative CL-STEM on mechanically exfoliated monolayers of TMDs enabled by encapsulation of a single TMD layer in hBN, which we apply to study effects of interface quality on opto-electronic properties. The hBN encapsulation enhances CL yield by increasing electron beam interaction volume, and reduces electron beam induced damage [19, 28]. We first focus on tungsten disulfide ($WS_2$) because of its high excitonic quantum efficiency at room temperature [29]. To demonstrate generality of our approach, we then extend it to weakly luminescent molybdenum disulfide ($MoS_2$), which displays only ~1% quantum efficiency at room temperature [29], as well as a representative alloy crystal (WSSe). We observe strong correlation between nanoscale luminescence variations and the interface quality between layers and monolayer folding. These measurements are the first direct nanoscale optical-structural characterization of an application-relevant interface (hBN-TMD-hBN), as well as representative monolayer alloy crystals. Finally, by measuring the effect of electron dose on the optical signal, we have laid out experimental conditions that minimize electron-beam effects to enable measurements of intrinsic properties of monolayer TMDs. These measurements show that CL-STEM is an important tool for characterizing TMDs and other atomically thin semiconductors, as well as application-relevant heterostructures.

To demonstrate feasibility of CL-STEM studies on single layer TMDs, we first started with exfoliated single layers of $WS_2$. Details of the experimental procedure are provided in the ESM; in short, the active material, monolayer $WS_2$, was first exfoliated from bulk crystals using the well-established scotch-tape method [1]. It was then sandwiched between two layers of hBN and this combined layered structure is henceforth referred to as the heterostructure. A modified dry transfer method was used to assemble the heterostructure on silicon dioxide/silicon, followed by annealing in an inert argon environment at 200°C for ~8 h [15]. The thickness of each hBN layer was >20 nm to ensure a sufficient interaction volume with the electron beam during CL measurements. The annealing process is critical to measure any CL-STEM on these samples, demonstrating the importance of a clean interface free of contaminants or trapped air. The annealing was followed by a wet transfer using polymethyl-methacrylate (PMMA) on a C-flat TEM grid, followed by cleaning of polymer with acetone and isopropanol (ESM, Figure S1).

White light bright field image of the heterostructure transferred to a TEM grid (Figure 1a) shows contrast variations expected from thickness variations of different layers. Before stacking, thickness of each layer was first estimated using optical contrast (high reflection indicates thicker hBN and low contrast indicates thin $WS_2$) and confirmed using atomic force microscopy for hBN, and high PL for monolayer $WS_2$ (ESM, Figure S2). A corresponding PL image of the heterostructure was measured using a 20× objective with green laser excitation (532 nm), and appropriate wavelength filtering to spectrally select the exciton emission (Figure 1b, see

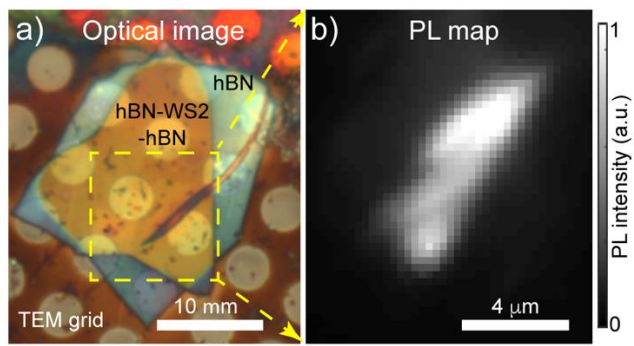

**Figure 1: a)** Optical image of hBN-$WS_2$-hBN heterostructure on a TEM grid. Green region corresponds to hBN only, whereas the sandwiched hBN-$WS_2$-hBN appears as the brown central region. The circular light brown regions are holes in the TEM grid. b) PL map of the outlined dashed box in (a).

ESM for image acquisition details). The PL map appears mostly homogeneous, but lacks the spatial resolution to probe nanoscale features. Further, it is not possible to measure interface quality *via* PL spectral line-width narrowing at room temperature, since the exciton is homogeneously broadened [9, 13, 30].

To achieve direct structure-property correlation with high spatial resolution, we next simultaneously collected STEM and CL images (Figure 2a). A JEOL 2011 TEM operating at an accelerating voltage of 80 kV was used to reduce knock-on damage in $WS_2$ [31]. STEM was operated in the dark field mode, where the contrast is a result of mass-thickness changes such that thicker (or higher atomic number) materials appear brighter. The STEM image (Figure 2b) clearly shows the location of the overlapping hBN layers and the sandwiched $WS_2$ monolayer (ESM, Figure S3 for annotated STEM image). Additional smaller contrast features, discussed in detail later, are due to folds in the $WS_2$ monolayer or PMMA residue. CL was measured through an integrated Gatan MONOCL3+ CL system equipped with a photo-multiplier tube detector [20]. Briefly, a focused electron beam (~ 1 nm) was incident on the heterostructure, generating electron hole pairs that, due to type 1 band-gap alignment, are funneled from hBN layers into the active TMD material (inset of Figure 2a). The light was sent through a spectrometer operated in the monochromatic mode to collect CL emission maps at specific wavelengths.

A representative CL spectrum is shown in the inset of Figure 2c as well as ESM Figure S4, and contains response from $WS_2$ as well as encapsulating hBN layers. The major spectral peak in CL is still excitonic, even though the excitation is non-resonant *via* high energy electrons and funneled electrons from encapsulating hBN layers. To measure the $WS_2$ optical response, we spectrally selected the $WS_2$ exciton resonance (616 nm) integrated over a spectral width of ~3 nm. The resulting CL monochromatic image (Figure 2c) shows a number of features that can be directly related to structural features in the STEM image. Specifically, a set of dark linear regions can be observed in the CL map across the monolayer, which can be directly correlated to linear folds that locally form multi-layers. Furthermore, a number of dark circular areas appear in the CL image, and we discuss these non-luminescent features in detail later.

We note that hBN layers not in direct contact with the monolayer $WS_2$ also display signal in the monochromatic CL image. To distinguish between these hBN and encapsulated monolayer $WS_2$ areas, monochromatic maps corresponding to the $WS_2$ exciton resonance at 616 nm as well as a non-resonant wavelength at 550 nm were taken sequentially (Figures 2c and d, respectively). The emission at 550 nm corresponds to a broad spectral background with contributions from partially oxidized hBN, trap states in hBN, as well as thermal radiation (centered around 400 nm) [32, 33]. On the other hand, emission at 616 nm has contributions from the same physical phenomena (which contribute to emission at 550 nm), and additionally from the comparatively narrow $WS_2$ exciton emission. Thus, a visual comparison of monochromatic images at 550 and 616 nm illustrates positions at which the sandwiched $WS_2$ monolayer luminesces.

To further visually isolate the $WS_2$ monolayer region and emphasize features due only to $WS_2$, in Figure 2e we plot CL intensity measured at 616 nm normalized to that measured at 550 nm. To confirm the spectral response and to characterize possible spectral shifts due to inhomogeneities (strain or doping), point spectra were taken at multiple positions (inset of Figure 2c). The two closely spaced positions (positions A and B) show a spectral shift ~ 10 nm, which could be related to strain or local doping; position B is supported on the TEM grid, suggesting strain-induced shift relative to position A [1, 34].

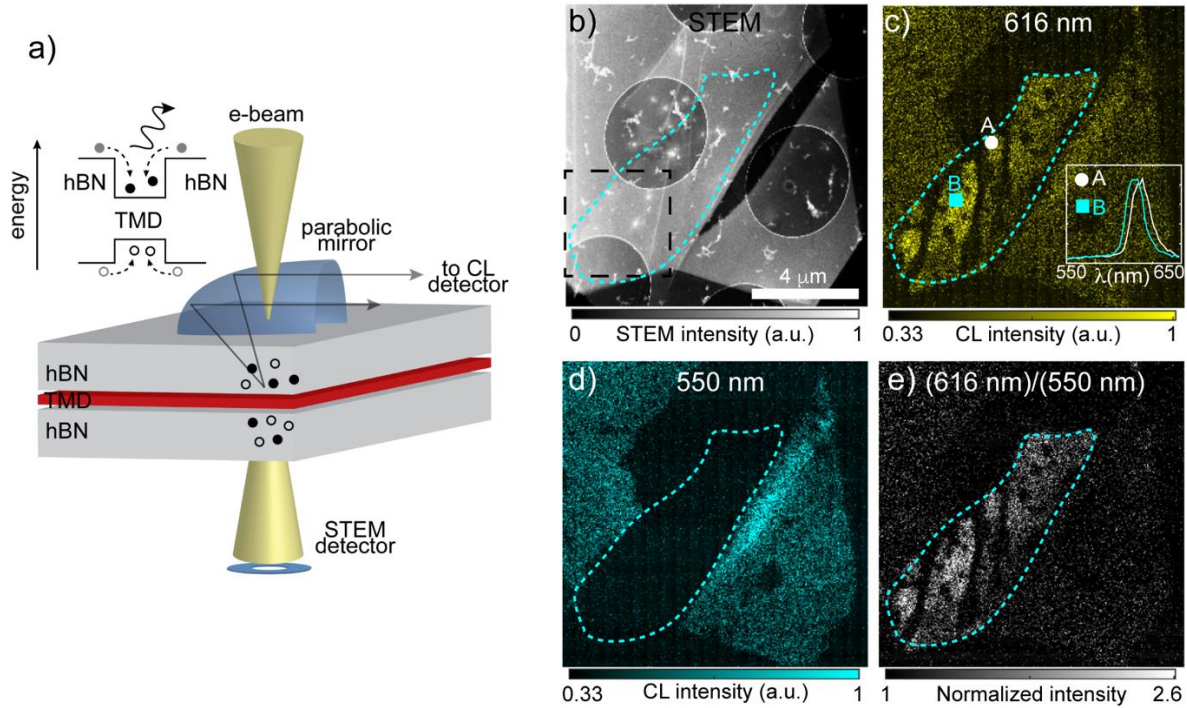

**Figure 2:** a) Schematic of the CL-STEM setup. Electrons and holes are generated in thick layers of hBN by a focused electron beam, and subsequently funnel into the TMD. The electron-hole pairs recombine radiatively, and subsequent light is collected by a parabolic mirror and directed toward a CL detector. Simultaneously with the light collection *via* CL, STEM detector provides structural information from the same sample region. The inset illustrates the corresponding type 1 band-gap alignment. b) STEM image of the hBN-$WS_2$-hBN heterostructure and c) the corresponding monochromatic CL image obtained at $WS_2$ exciton resonance, 616 nm. $WS_2$ monolayer region is indicated by the blue dotted lines. The black dashed box in (b) is measured in Figure 3. Inset: CL spectrum of two positions indicated by white circle (A) and yellow box (B). d) Monochromatic CL image at 550 nm. e) Image obtained after normalizing monochromatic data at 616 nm (exciton resonance) with data at 550 nm (broad background) to emphasize $WS_2$ region. Scale bar is identical for (b-e). All intensities are plotted on a linear scale. Maximum values in each figure were chosen to enhance contrast.

These results illustrate that emission from a single $WS_2$ layer can be successfully obtained using CL-STEM. We now focus on nanoscale heterogeneities in the CL images to uncover their structural and physical origin. A high magnification (50 k×) monochromatic CL image at the exciton resonance (616 nm) is shown in Figure 3b along with the corresponding STEM image in Figure 3a. Two general types of dark features – *i.e.* areas with reduced or vanishing $WS_2$ luminescence – can be observed in Figure 3(b) with distinct appearance and underlying cause: (1) elongated bands running from top to bottom of the figure and (2) circular regions that do not show any specific spatial arrangement. The dark bands (in CL) correspond to bright lines in the STEM image and therefore can be assigned to higher mass, *i.e.* few layers of $WS_2$ that are formed by folding of a single sheet, and consequent transition from a direct to indirect bandgap. On the other hand, dark circular areas do not show any apparent STEM contrast compared to surrounding luminescent areas, implying no thickness or compositional change. It has been shown that stacking or sandwiching of atomically thin layers can lead to formation of nanoscale "bubbles" due to localized trapped air or contaminants [35]. These are unlikely to produce a significant STEM contrast, but we suggest are significant enough to reduce interlayer coupling between hBN and $WS_2$ and dramatically reduce CL intensity. Our results show that CL-STEM acts as a high-resolution diagnostics tool that provides information about these inhomogeneities and their role on interlayer coupling, even in the absence of any structural signature in STEM.

We next comment on the spatial resolution of CL-STEM in assessing the interlayer coupling. The high spatial resolution of the measurement is exemplified by a CL image (Figure 3c) obtained at even higher magnification of 100 k× from another encapsulated hBN-$WS_2$-hBN heterostructure (Sample-B). To characterize the spatial resolution, we take a line intensity profile along a dark feature (attributed to air bubble), and measure a full width half maximum of ~ 100 nm along the dark feature. We note that the spatial resolution is likely limited by signal to noise ratio of our detection system as well as lack of smaller features in this monolayer sample [24]. Factors contributing to signal to noise ratio are background signals from CL (stray electrons hitting TEM grid, mirrors etc.), detector thermal noise and dark counts. Further increase in the signal could be achieved by using thicker layers of hBN, but this may cause degradation of the spatial resolution due to lateral diffusion of carriers within hBN before funneling into $WS_2$. Hence, an optimum thickness of hBN may be required, coupled with high efficiency detection. Alternatively, other encapsulating materials including higher quality hBN, plasmonic materials and scintillating sources could be explored to enhance light emission [36, 37]. After optimizing for hBN thickness and interface quality, evaluating other encapsulating materials, as well as improving detectors, measurements of atomic structure and single point defects could be explored. In the ESM (Figure S9, S10 and S11), we also discuss spatial resolution differences between CL-SEM and CL-STEM. We find that CL-STEM offers higher spatial resolution, and direct structure-optical correspondence. We note that understanding of the CL resolution also enabled us to distinguish between (multi)layer folding and termination edges, from which we show that folding has longer-range effects on the interlayer coupling (see ESM, Figure S5).

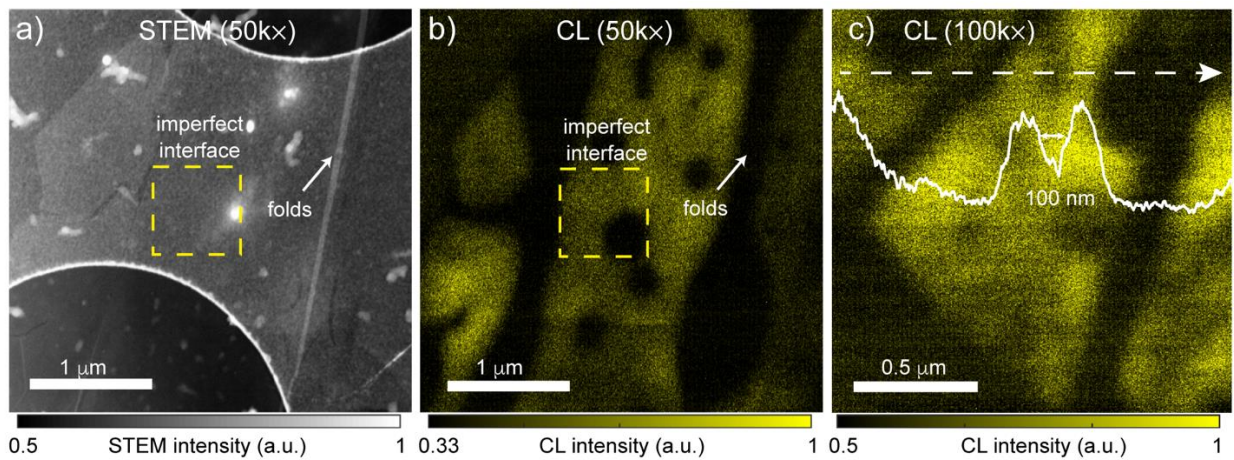

**Figure 3:** (a) STEM image and (b) corresponding monochromatic CL image at exciton resonance (616 nm) at 50 k× magnification. Scale bars in both cases are 1 $\mu$m. The exact measurement area is indicated in Figure 2b with a black dashed box. Localized bright spots in STEM correspond to positions of earlier point spectra measurements, resulting in carbon contamination. c) Monochromatic CL image from another encapsulated hBN-WS$_2$-hBN heterostructure (Sample-B) at high magnification (100 k×). Spatial resolution is measured by taking a line intensity profile along the white dashed line crossing a dark spot (attributed to an air bubble). Line profile (solid white line) of CL intensity indicates spatial resolution of ~ 100 nm. Maximum values in each figure were chosen to enhance contrast.

After establishing CL-STEM as a method to directly visualize interlayer coupling in WS$_2$ as a model system, we now extend the technique to other 2D materials, specifically monolayer MoS$_2$ and an alloy monolayer WS$_{0.8}$Se$_{1.2}$. Remarkably, using our approach we could measure significant CL even for a low quantum efficiency material MoS$_2$. Like in the case of WS$_2$, CL maps of hBN-sandwiched monolayer MoS$_2$ shows heterogeneous features (Figure 4a). Dark features in CL images can be directly correlated to bright features in STEM (Figure 4b), which are most likely related to more significant interlayer contamination. Monolayer WSSe alloy crystals show qualitatively similar heterogeneities in CL-STEM. The STEM image (Figure 4d) shows increased intensity at areas which are dark in the corresponding CL image (Figure 4c). These areas are again attributed to larger contamination. We also measured point spectra at different positions in the alloy crystal, and do not observe any significant spectral shifts (not shown here). Thus, these alloy crystals are not observed to have gross compositional fluctuations at the resolution of our measurement. Together, these results show that CL-STEM measurements enable evaluation of the interface quality in a range of TMDs, and are thus general in nature.

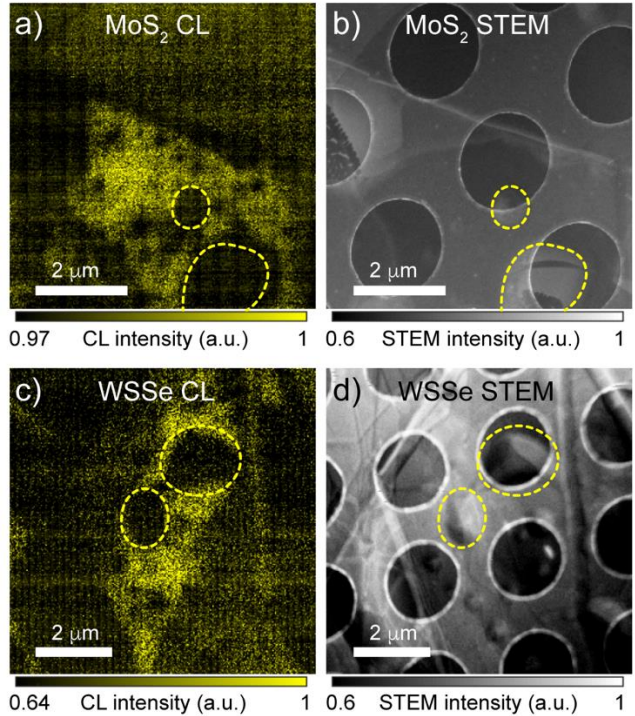

**Figure 4:** (a) Monochromatic CL image (at exciton resonance, 646 nm) and (b) corresponding STEM image of a sandwiched MoS$_2$ monolayer. (c) Monochromatic CL image (at exciton resonance, 720 nm nm) and (d) corresponding STEM image of a sandwiched WSSe monolayer. Dark CL regions for both samples are indicated by dashed lines. Maximum values in each figure were chosen to enhance contrast.

Finally, to investigate the possible role of the electron beam on CL – and to ascertain that CL inhomogeneities discussed above are not caused by possible structural damage by electron beam – we characterized evolution of the CL signal as a function of the electron beam exposure. When electron beam is focused on a specific position, *e.g.* to obtain point CL spectra, we observed reduction of ~ 50% in CL intensity when comparing subsequent point spectra acquired for 70 s (1$^{st}$ and 2$^{nd}$ scan in Figure 5a), and reduction by 5-10% in subsequent scans (compare 7$^{th}$ scan with 2$^{nd}$ scan). At the accelerating voltages used in the experiment (80 keV), direct knock-on damage is reduced, but not completely prevented [38]. The sulfur vacancy damage threshold for encapsulated samples is expected to be higher compared to a bare monolayer as the encapsulation provides a protective barrier [28], and prevents escape of the removed sulfur into the vacuum [39]. The self-limiting electron beam induced damage is suggested to be a result of direct knock-on at a small area, whereas the luminescence is expected to result from a larger area due to lateral diffusion of carriers in surrounding hBN layers. Although damage can occur in surrounding hBN layers [40], it is expected to be limited due to their thickness.

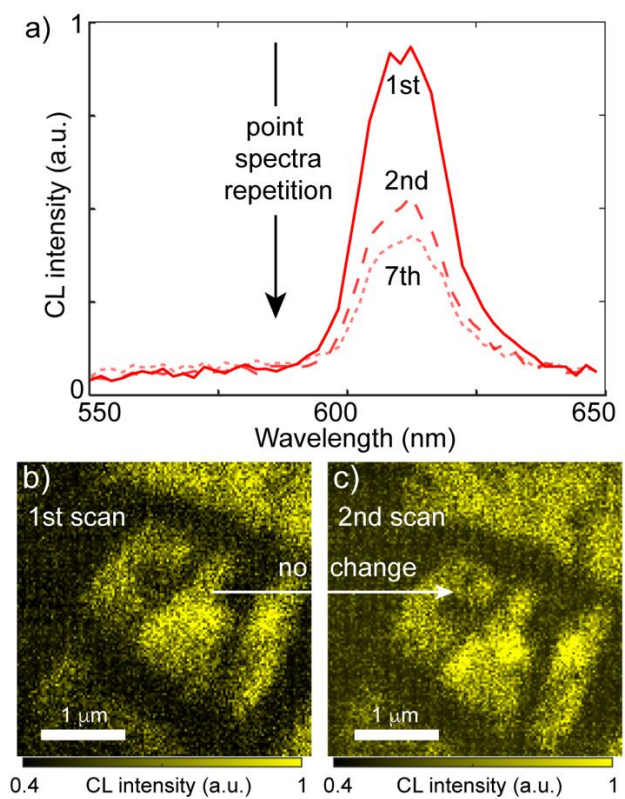

**Figure 5**: a) Sequential (1st, 2nd, and 7th) point spectra of $WS_2$ each acquired for 70 s, illustrating beam-induced reduction of CL intensity. b), c) Comparison of monochromatic images for (b) first and (c) subsequent data collection on encapsulated $WS_2$ Sample-B. Maximum values in each figure were chosen to enhance contrast.

In contrast to the point spectra, monochromatic mapping requires significantly shorter acquisition time of 1.4 ms/pixel with the same current density (50 k× lower dose), for which no observable change in the intensity or in the spatial extent of the luminescence was observed (Sample-B, Figures 5b and 5c and the ESM S6). Therefore, we conclude that the CL inhomogeneities measured in regions in which no significant e-beam dwelling was applied (Figures 2-4) are intrinsic to the heterostructure, and not electron-beam induced artifacts. Thus, measuring monochromatic images offers a suitable way to characterize nanoscale areas without causing damage.

In conclusion, we have utilized CL-STEM to measure nanoscale heterogeneities in monolayer TMDs ($WS_2$, $MoS_2$, WSSe) encapsulated in dielectric layers (hBN). We directly measured the effect of an imperfect interface – resulting from contaminants and layer folding – on the luminescence, resulting in the first direct nanoscale optical-structural characterization of the interface. As these nanoscale inhomogeneities are a general feature of single phase and multi-phase heterostructures, it is important to understand their role for enabling higher quality opto-electronic devices. Further, we observed that even excitation by high energy electrons (80 kV) in a TEM primarily results in ground state exciton CL, allowing direct comparison with PL measurements, but with higher spatial resolution. We quantified the effect of electron beam irradiation on the CL magnitude, and demonstrated a suitable imaging condition with reduced damage to the sample. These first CL-STEM measurements on monolayer TMDs should pave the way forward for further direct nanoscale structure-optical measurements in application relevant multi-phase materials. By enhancing signal to noise by using thicker hBN and improved detectors, direct defect mapping may be possible in future measurements.

ASSOCIATED CONTENT

**Electronic Supplementary Material (ESM)**. CL-STEM experimental conditions, discussion on electron-hole pairs generated in hBN layer diffusing to WS$_2$, preparation of heterostructure, PL and atomic force microscopy image measurements, annotated STEM image, CL Spectrum, spatial extent of folded layers measured via STEM and CL, sequential electron beam monochromatic images, effect of encapsulation and annealing on CL, differences between CL-SEM and CL-STEM. The files are available free of charge.

ACKNOWLEDGMENTS

This work was supported by the Singapore-MIT Alliance for Research and Technology and the National Research Foundation, Prime Minister's Office, Singapore. The authors acknowledge helpful discussions with Yuan Cao and Kha Tran. This work made use of the MRSEC Shared Experimental Facilities at MIT, supported by the National Science Foundation under award number DMR-1419807. Any opinion, findings, and conclusions or recommendations expressed in this material are those of the authors and do not necessarily reflect the views of the National Science Foundation.

# Direct Optical-Structure Correlation in Atomically Thin Dichalcogenides and Heterostructures

*Akshay Singh, Hae Yeon Lee, Silvija Gradečak\**

Department of Materials Science and Engineering, Massachusetts Institute of Technology, 77 Massachusetts Ave, Cambridge, MA 02141, USA

**CL-STEM experimental details**

We utilized a JEOL-2011 transmission electron microscope (TEM) in dark field scanning transmission electron microscope (STEM) mode, at an accelerating voltage of 80 kV. The microscope is equipped with the Gatan MONOCL3+ cathodoluminescence (CL) collection system for simultaneous STEM and CL measurements. The focused electron beam (~ 1 nm in diameter) is incident on the sample and creates electron-hole pairs, which recombine either radiatively or nonradiatively. In the case of radiative recombination, the resulting light is collected by a parabolic mirror positioned between the TEM grid and the TEM pole piece. The signal is spectrally dispersed using a monochromator, and subsequently sent to a photomultiplier tube (PMT) for detection. Monochromatic images are taken either with 256×256 pixels or 512×512 pixels, with integration time of 0.7 ms or 1.4 ms (per pixel), respectively. Point spectra are measured over a spectral range of ~ 100 nm with step ~ 2 nm and an integration time of ~ 1.4 s per step. Simultaneously, the dark field STEM detector collects scattered electrons resulting from incoherent interactions with the sample, with the contrast directly proportional to mass (or thickness). The incident electron beam current is ~ 1 nA, and the convergence angle is ~ 3 mrad.

**Light generation in monolayer WS$_2$**

The WS$_2$ layer is sandwiched between two thick layers (> 20 nm) of hBN. Electron-hole pairs are generated in hBN and monolayer WS$_2$, with the number of generated electron-hole pairs approximately proportional to the layer thickness. hBN has a bandgap of ~ 6 eV and WS$_2$ has a bandgap of ~ 2.3 eV, and together they form type 1 band alignment. Hence, most of electron-beam generated electrons and holes diffuse into the WS$_2$ layer before recombining (see Figure 2 of the main text). Subsequently, electrons and holes recombine in WS$_2$ and generate light corresponding to the A-exciton of WS$_2$.

**Preparation of the hBN-encapsulated WS$_2$ heterostructure**

To prepare the hBN-encapsulated WS$_2$ heterostructure, we utilize a modified dry transfer method (Figure S1). WS$_2$ and hBN crystals were purchased from 2dsemiconductors and HQGraphene, respectively. Firstly, all component layers are individually mechanically exfoliated on silicon dioxide/silicon using scotch-tape method[1]. A PDMS (polydimethylsiloxane) mask (~ 4 mm thick) is prepared by using 20:1 ratio of Sylgard 184 pre-polymer to curing agent, and cured by keeping at ambient conditions for ~ 24 h. The PDMS is oxygen plasma treated (18 W) for 5 min, with subsequent spin coating of polypropylene carbonate (15% PPC in anisole) at ~ 3000 rpm. Plasma treatment improves adhesion of PPC onto PDMS and reduces de-wetting. The PPC-PDMS mask is heat-treated at 160° C for ~ 10 min to remove the solvent. This mask is mounted upside down on a transfer arm, and used to perform the transfer using a modified process from Ref[2]. Briefly we describe the procedure below,

1. The PPC-PDMS mask contacts the top hBN (kept on a $SiO_2$/Si substrate) at ~ 50° C, then temperature is increased to ~ 70° C (held for ~ 1 min), and then decreased to 50° C (held for ~ 1 min). By moving the mask away from the substrate, hBN is picked up.

2. The top hBN is dropped onto monolayer $WS_2$ by contacting the hBN-PPC-PDMS mask to the $WS_2$ at ~ 75° C, and then moving slowly away from the substrate. After the top hBN drops on $WS_2$, we heat treat the substrate at 75° C for ~ 5 min to improve adhesion between $WS_2$ and top hBN.

3. hBN-$WS_2$ is then picked up by PPC-PDMS by repeating step 1.

4. hBN-$WS_2$ is dropped on bottom hBN at ~ 100° C. This whole stack is then heated at ~ 120° C for 10 min to improve adhesion. A quick solvent cleaning step follows (10 s acetone, 10 s isopropanol) to remove minor PPC residue.

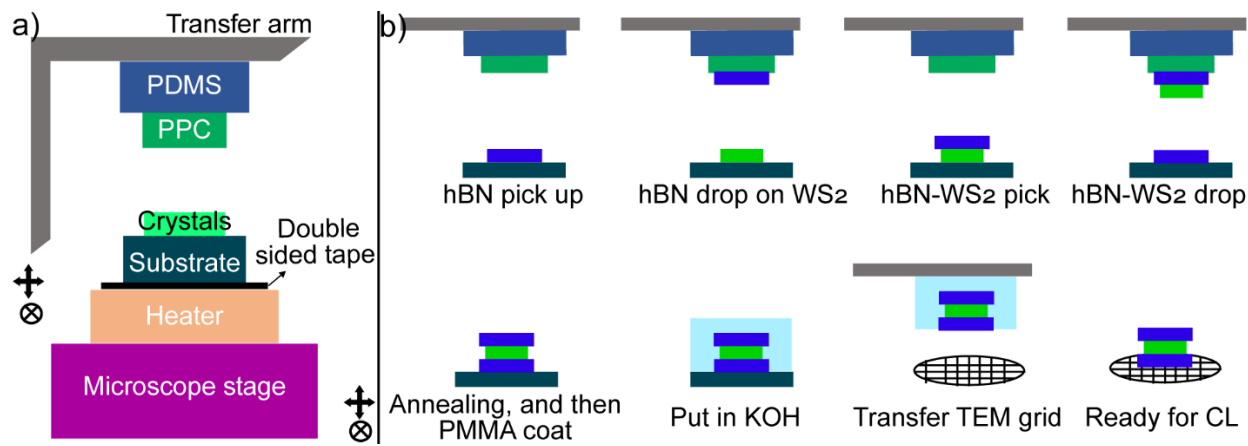

**Figure S1:** a) Schematic of transfer setup. b) Schematic of steps used in the transfer process.

5. Annealing at 200° C in a pure argon atmosphere is performed for ~ 8 h. This markedly improves the quality of interface between hBN and $WS_2$. PMMA (495K, A4 Microchem) is spin-coated on the heterostructure-SiO$_2$/Si stack, and clean room blue tape with a suitable opening (window) is applied.

6. The PMMA coated substrate is kept in 1 Molar KOH (Potassium Hydroxide) at room temperature for ~ 8 h, resulting in etching of SiO$_2$, and release of crystals.

7. The blue tape supported PMMA is then transferred to a C-flat grid (Protochips Inc., with either 200 mesh, 2 $\mu$m holes; or 400 mesh with 4 $\mu$m holes).

8. A quick solvent cleaning step (20 s acetone, 20 s IPA) finally prepares the crystals for CL-STEM.

**Photoluminescence (PL) and atomic force microscopy image (AFM) measurements**

Monolayer WS$_2$ is identified by optical contrast, and confirmed by PL measurements. The PL measurements are performed by using a green laser (532 nm) focused with a 20× or a 60× microscope objective, ~ 20 $\mu$m spot size, and power ~ 4 mW. The PL signal is sent to the spectrometer via a long-pass filter (to filter excitation laser). After carefully eliminating contributions from scattered laser and room light, we use the spectrometer in non-dispersive mode to measure a luminescence image (as seen in Figure 1(b) of the main text). We can also use in-built mechanical slits to narrow down to a small area, and measure the spectrum, shown in Figure S2(a). The spectral peak is 620 nm, similar to CL spectral peak. Further, we show an AFM line-scan of bottom hBN in Figure S2(b). The thickness of the bottom hBN is ~ 60 nm, whereas the thickness of top hBN is ~ 30 nm (not shown here).

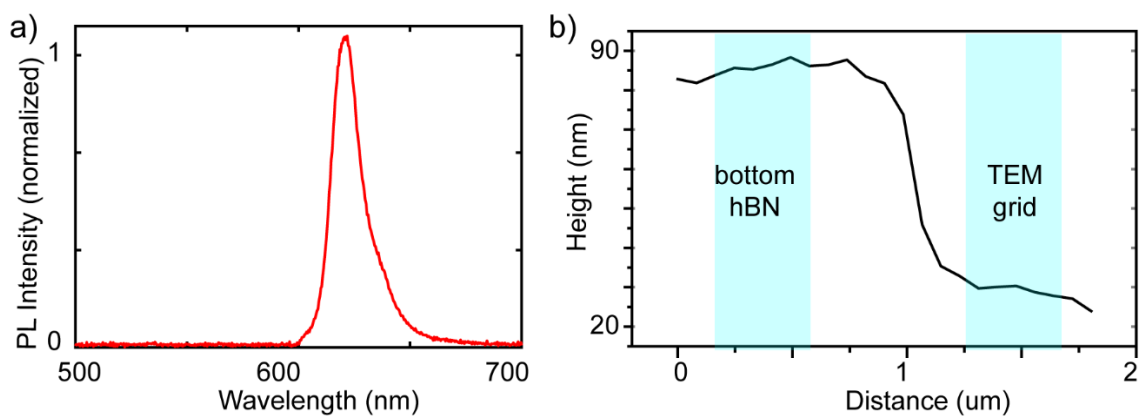

**Figure S2:** a) PL spectrum of monolayer WS$_2$. The luminescence peak is at 620 nm. b) AFM image of bottom hBN (of the heterostructure) mounted on a TEM grid.

**Annotated STEM image**

A dark-field STEM image of the area corresponding to Figure 1 of the main text is shown in Figure S3. We have indicated different areas with corresponding different STEM contrast. A clear distinction can be made between overlap regions of top and bottom hBN, and only bottom hBN. We also observe localized PMMA residue, resulting from incomplete removal by solvent cleaning process. The monolayer has a small contrast in this image, and is indicated by blue dotted lines. Monolayer folds are also shown, where the monolayer has folded onto itself resulting from imperfect transfer. Holes in the C-flat TEM grid (20 nm carbon thickness) can be clearly seen.

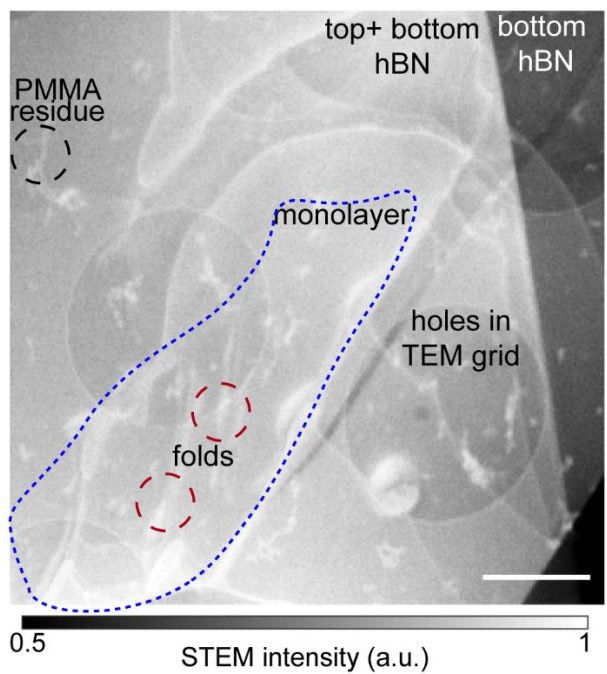

**Figure S3:** Annotated STEM image for region of interest (corresponding to Figure 1 of the main text). Changes in STEM contrast are attributed to PMMA residue, hBN thickness changes, monolayer (blue dotted area), monolayer folding and holes in the C-flat TEM grid. Scale bar is 2 $\mu$m.

**CL spectrum of the encapsulated WS$_2$**

To show the background (from hBN, as explained in main text) and excitonic contributions (WS$_2$ A-exciton) to CL spectrum, we show spectrum from position B (indicated in Figure 2c of the main text). As we discuss in the main text, the major contribution to CL on an encapsulated monolayer is of excitonic origin.

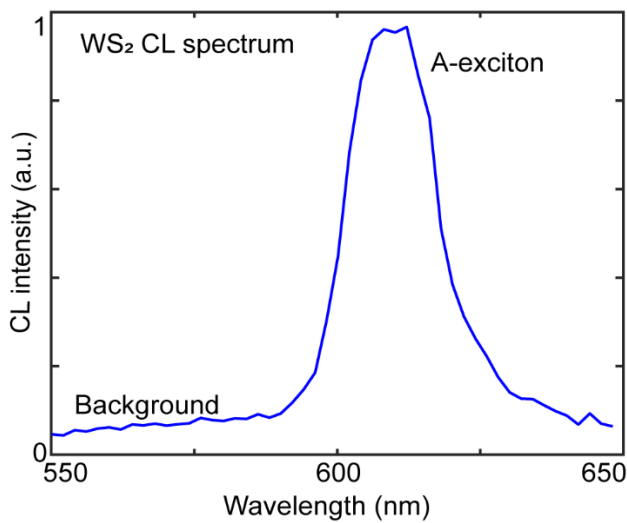

**Figure S4:** CL spectrum on encapsulated WS$_2$ monolayer (position B indicated in Figure 2d in the main text). The excitonic resonance at this position is measured at ~ 610 nm.

**Spatial extent of folded layers measured via STEM and CL**

Figure S5 shows a high resolution (50 k×) monochromatic CL image at WS$_2$ exciton resonance (616 nm), and corresponding STEM image for another hBN-WS$_2$-hBN heterostructure. The STEM image shows two types of edges in encapsulated WS$_2$ indicated by arrows. The edge along the solid arrow shows enhanced STEM contrast and is attributed to layer folding during the transfer process. A STEM contrast gradient transverse to this edge is observed, suggesting trapped contamination. In contrast, the other edge (along the dotted arrow) shows a sharp STEM contrast, therefore indicating a monolayer terminating edge rather than a folded area. The corresponding CL image indicates a wider dark region corresponding to the folded edge (solid line) compared to the monolayer edge (dotted line). Specifically, the CL intensity starts to decrease significantly around 200 nm from the folded edge, whereas CL intensity decreases in an area much closer to the monolayer edge. Thus, comparison of the two edges clearly shows the large length scale impact of imperfect edges, and indicates the importance of a perfect interface.

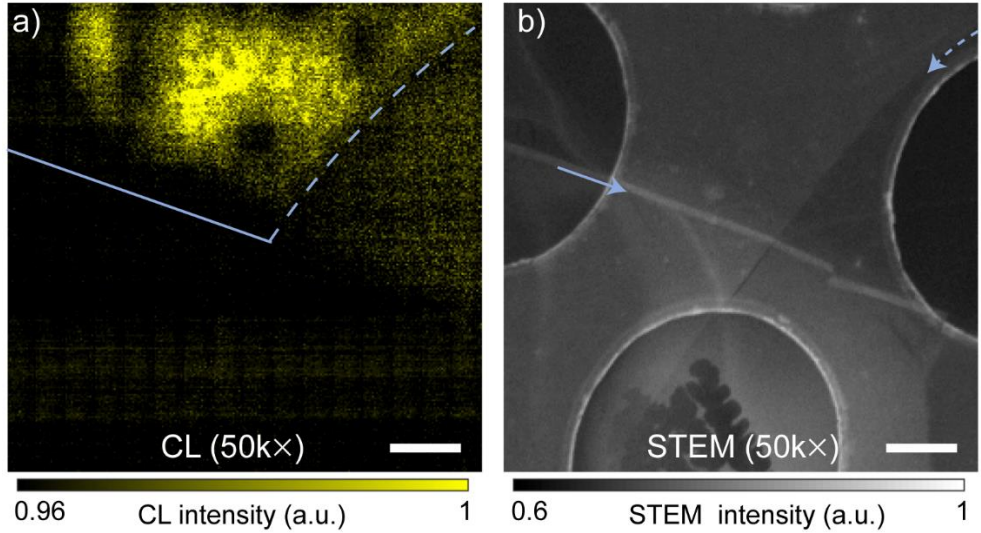

**Figure S5:** CL monochromatic image at $WS_2$ exciton resonance (616 nm), and corresponding STEM image for another hBN-$WS_2$-hBN heterostructure. Dotted (solid) line in a) corresponds to the monolayer (folded layer) edge, again shown in b) as dotted (solid) arrow. Scale bars are 500 nm.

**Sequential electron beam monochromatic images**

As mentioned in the main text, the damage from the electron beam is minimal when monochromatic images are taken. Here we show sequential monochromatic images (256×256 pixels, 0.7 ms per pixel), which show almost no change in the CL signal.

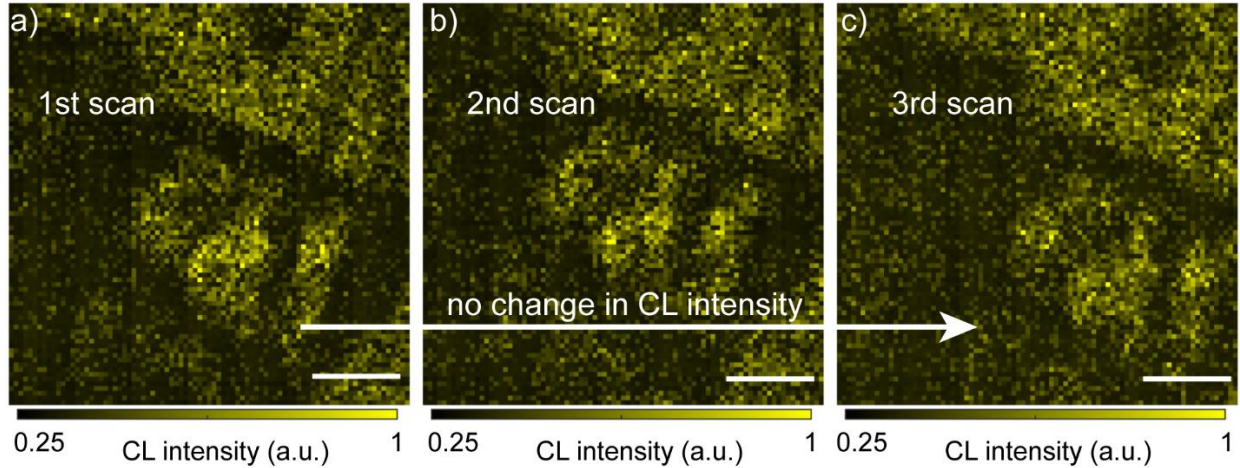

**Figure S6:** Sequential monochromatic images at the WS$_2$ exciton resonance, 616 nm (in the order a, b, c) for Sample-B. All scale bars are 500 nm. There is a small change in the sample position between measurements due to a sample drift.

**Effect of encapsulation and annealing on CL**

As we have mentioned in the manuscript, without the hBN encapsulation, negligible CL is measured for any sample. Further, the lack of luminescence in the area of imperfect interfaces illustrates the effect of improper hBN encapsulation. As an example experiment, we show CL in a bare (not encapsulated) monolayer WS$_2$. In the CL image, the monolayer area is barely distinguishable from the TEM grid and background noise. The mechanism of the enhancement is detailed in the manuscript and is due to higher number of beam-induced carriers in thicker encapsulated samples that subsequently recombine in a TMD monolayer.

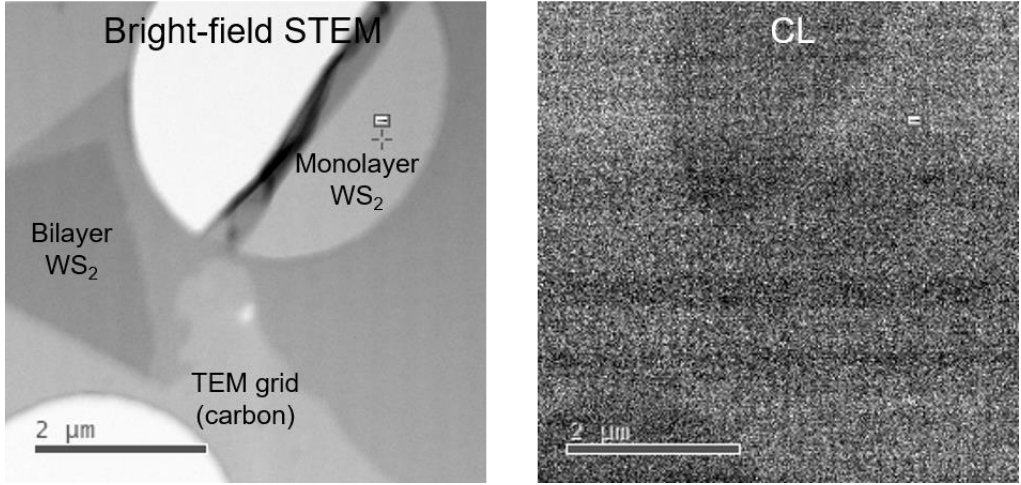

**Figure S7:** Bright field STEM and corresponding CL image on un-encapsulated WS2 monolayer sample. Negligible CL is detected.

Another important aspect of sample preparation is annealing of the heterostructure. Our process involves first annealing the heterostructure sample, and then transfer to TEM grid. This order preserves the integrity of the carbon on the TEM grid. However, as proof of concept, we show CL-SEM before annealing on a heterostructure sample. For comparison, another sample after annealing is also shown. These samples are representative of the effects of annealing on a large (>10) number of samples.

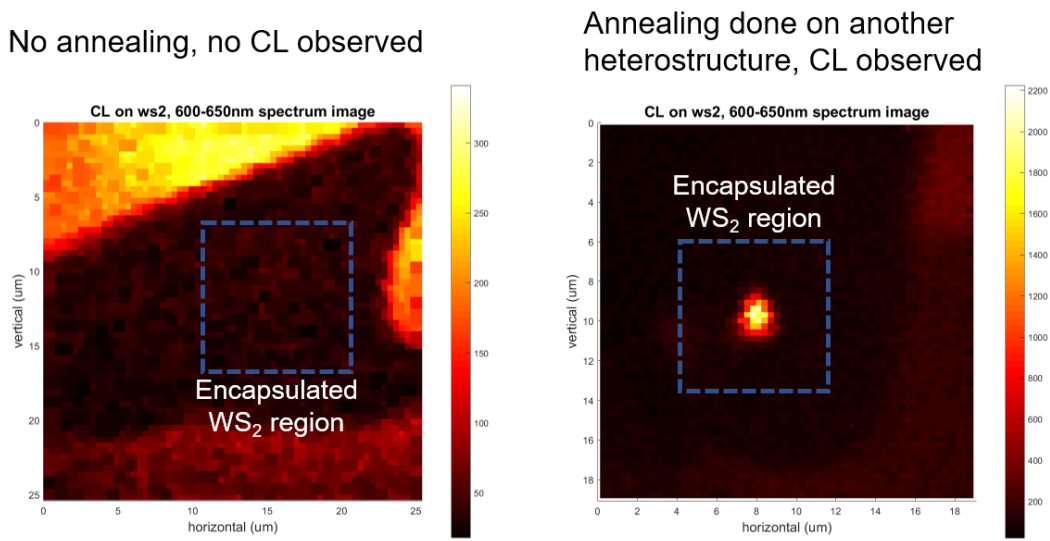

**Figure S8:** Effect of annealing on two separate heterostructures. Left- spectrally integrated ($WS_2$ exciton resonance) CL-SEM image on un-annealed heterostructure. Right- spectrally integrated CL-SEM image on an annealed heterostructure, showing CL corresponding to encapsulated $WS_2$.

**Differences between CL-SEM and CL-STEM**

Both CL-SEM and CL-STEM are well established methods. However, there are some clear advantages of STEM-CL vs. SEM-CL namely (i) improved spatial resolution in both the optical and structural information and (ii) direct structure-property correlation.

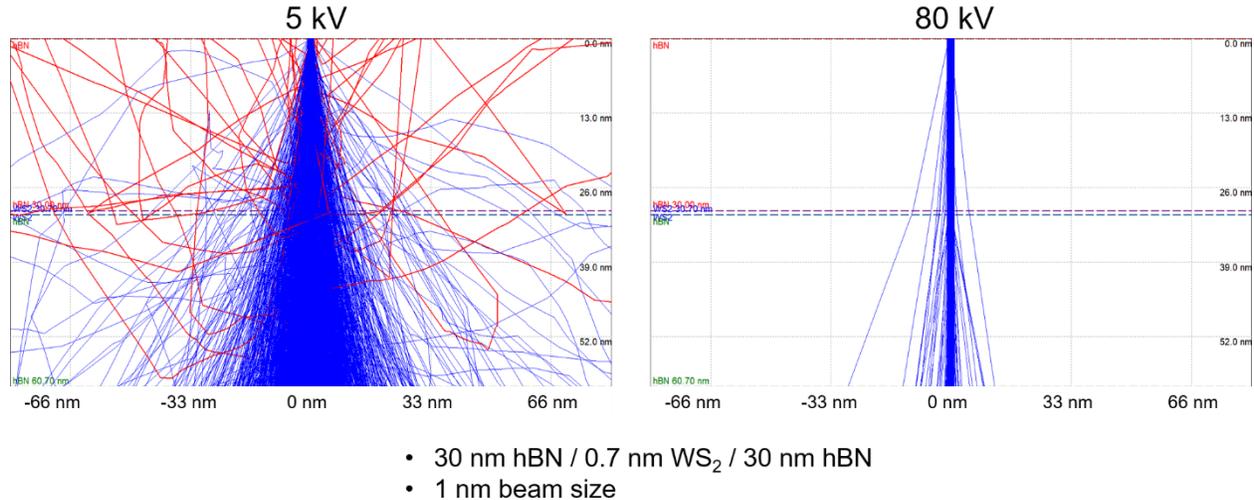

- 30 nm hBN / 0.7 nm $WS_2$ / 30 nm hBN
- 1 nm beam size

**Figure S9**: Monte-Carlo simulations performed on a typical hBN-$WS_2$-hBN heterostructure. 5 kV beam clearly broadens (higher interaction volume) much more than 80 kV, indicating higher spatial resolution for 80 kV electron beam.

Due to higher electron energy in (S)TEM compared to SEM, interaction volume is reduced, as illustrated by Monte-Carlo simulations (Figure S9, performed using CASINO software [3]). For a typical heterostructure investigated in our work, 1 nm beam probe will broaden to ~ 30 nm, whereas in STEM this beam broadening will be minimized, which improves CL resolution. An example of higher spatial resolution for CL-STEM is illustrated in figure S10, where a hBN-$WS_2$- hBN heterostructure is measured via SEM and STEM. The CL-STEM image shows far more features, which indicates higher spatial resolution.

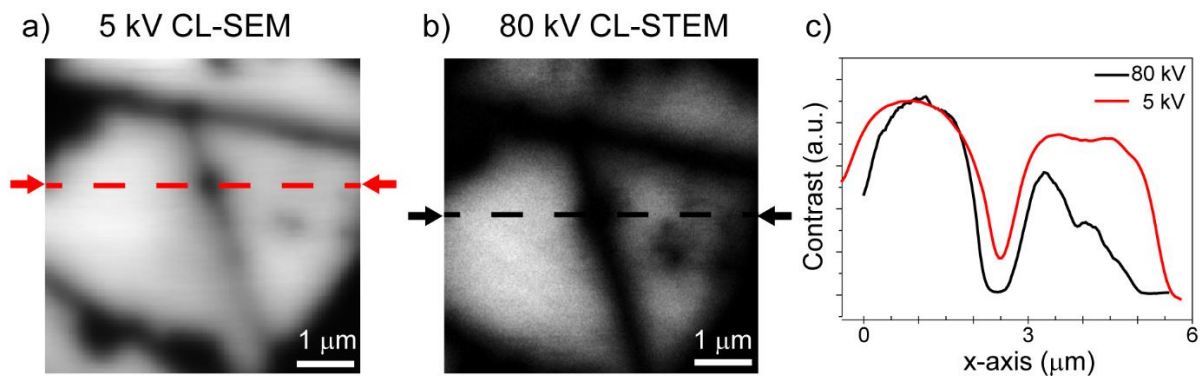

**Figure S10**: Comparison of CL monochromatic images for accelerating voltage a) 5 kV (SEM), b) 80 kV (STEM). The heterostructure is hBN-$WS_2$-hBN with similar thicknesses as in main text. Line cuts in both figures across dark features are compared in (c). CL measured via 80kV can detect many spatial features, indicating higher spatial resolution.

In STEM-CL, structural characterization with nanoscale resolution (STEM) is recorded simultaneously with the CL signal, which provides direct structure-property correlation and high contrast for the monolayer compared to hBN encapsulating layers. However, in an SEM, relying on secondary electron contrast (or back-scattered electron contrast) allows only limited contrast for the monolayer, compared to surrounding encapsulating layers (Figure S11).

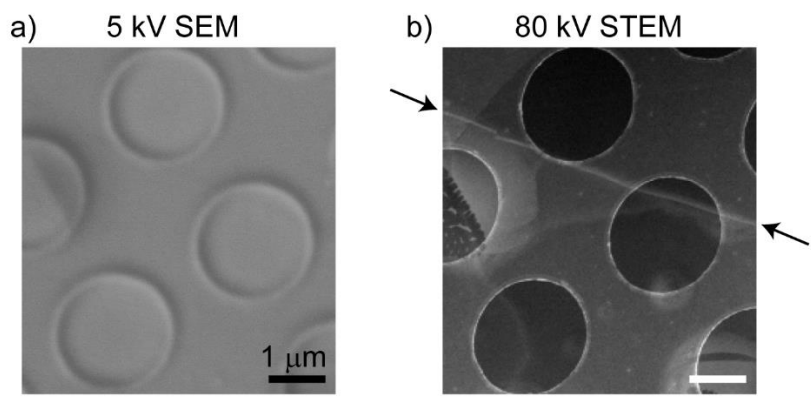

**Figure S11**: a) Secondary electron contrast for heterostructure (same as in Figure S5) measured by SEM. b) STEM contrast for same heterostructure clearly identifies monolayer and layer folding.